\documentclass[aps,prb,preprint,superscriptaddress]{revtex4}

\usepackage{pkgs_r}
\usepackage{mcrs}

\begin{document}

\title{Knowledge-Transfer-Based Cost-Effective Search for Interface Structures: \\
A Case Study on fcc-Al [110] Tilt Grain Boundary}


\author{Tomohiro~Yonezu}
\affiliation{Department of Computer Science, Nagoya Institute of Technology, Gokiso, Showa, Nagoya, Aichi 466-8555, Japan}

\author{Tomoyuki Tamura}
\email{tamura.tomoyuki@nitech.ac.jp}
\affiliation{Department of Physical Science and Engineering, Nagoya Institute of Technology, Gokiso, Showa, Nagoya, Aichi 466-8555, Japan}
\affiliation{Center for Materials research by Information Integration, National Institute for Materials Science, Tsukuba 305-0047, Japan}

\author{Ichiro Takeuchi}
\affiliation{Department of Computer Science, Nagoya Institute of Technology, Gokiso, Showa, Nagoya, Aichi 466-8555, Japan}
\affiliation{Center for Materials research by Information Integration, National Institute for Materials Science, Tsukuba 305-0047, Japan}

\author{Masayuki Karasuyama}
\email{karasuyama@nitech.ac.jp}
\affiliation{Department of Computer Science, Nagoya Institute of Technology, Gokiso, Showa, Nagoya, Aichi 466-8555, Japan}
\affiliation{Center for Materials research by Information Integration, National Institute for Materials Science, Tsukuba 305-0047, Japan}
\affiliation{PRESTO, Japan Science and Technological Agency, 4-1-8 Honcho, Kawaguchi, Saitama 332-0012, Japan}


\begin{abstract}
 Determining the atomic configuration of an interface is one of the most important issues in materials science research.
 Although theoretical simulations are effective tools, an exhaustive search is computationally prohibitive due to the high degrees of freedom of the interface structure.
 %
In the interface structure search, multiple energy surfaces created by a variety of orientation angles need to be explored, and the necessary computational costs for different angles vary substantially owing to significant variations in the supercell sizes.
 In this paper, we introduce two machine-learning concepts, called transfer learning and cost-sensitive search, to the interface-structure search.
 As a case study, we demonstrate the effectiveness of our method, called cost-sensitive multi-task Bayesian optimization (CMB), using the fcc-Al [110] tilt grain boundary. 
 Four microscopic parameters, the three-dimensional rigid body translation, and the number of atomic columns, are optimized by transferring knowledge of energy surfaces among different orientation angles. 
 We show that transferring knowledge of different energy surfaces can accelerate the structure search, and that considering the cost variations further improves the total efficiency.
\end{abstract}

\maketitle


\clearpage

\section*{Introduction}
\label{sec:introduction}

A \emph{grain boundary} (GB) is the interface between two grains or crystals
in a polycrystalline material, and has an
atomic configuration significantly different 
from that of a single crystal.
Since this 
results in
peculiar mechanical and electrical properties of materials, one of the most important issues in materials research is determining the atomic configuration of an interface. 
Experimental observations, such as the atomic-resolution transmission electron microscope (TEM) observations\cite{Haider1998} and theoretical simulations, such as first-principles calculations based on the density functional theory and static lattice calculations with empirical potentials, have been extensively performed to investigate interface structures\cite{Alfthan2010,Ikuhara2011}.

The macroscopic GB geometry is defined using five degrees of freedom (DOF) that fully describe the crystallographic orientation of one grain relative to the other (3 DOF) and the orientation of the boundary relative to one of the grains, i.e., the GB plane (2 DOF). 
Besides these five macroscopic DOF, three other microscopic parameters exist for relative rigid body translation (RBT) of one grain to the other parallel and perpendicular to the GB plane. 
It has been indicated that the most important parameter in determining the GB energy is the excess boundary volume \cite{Wolf1989}, which is related to the RBT perpendicular to the boundary. 
Closely packed boundaries that have a local atomic density similar to that in the bulk will have low energies. 
Thus, it is important to determine both the RBT and the number of atomic columns at the boundary \cite{Muller1999}. 
%
These microscopic parameters are established based on energetic considerations and cannot be selected arbitrarily, and atomistic simulations are widely used to obtain stable GB structures. 
%
To understand the whole nature of GBs, the stable interface structures for each rotation angle and rotation axis need to be determined.
%
A straightforward manner is optimizing all possible candidates of GB models, thereby determining the lowest-energy configuration. 
However, determining the stable structures of GBs needs large-space searching due to the huge geometric DOF.
Although some databases of GB structures are available \cite{Olmsted2009,Erwin2012,Banadaki2016}, they contain only a limited number of systems because of considerable computational costs of simulations.
Therefore, developing efficient approaches to determining the interface structure without searching for all possible candidates is strongly demanded.

In recent years, materials-informatics techniques based on \emph{machine learning} have been introduced as an efficient way for  data-driven material discovery and analysis \cite{Rodgers2006}.
%
For the structure search, which is our main focus in this study, a machine learning technique called \emph{Bayesian optimization} \cite{Shahriari16} has proven to be useful mainly in the application to determine stable bulk structures \cite{Seko2015,Ueno2016}.
Bayesian optimization iteratively \emph{samples} a candidate structure predicted by a probabilistic model that is statistically constructed by using already sampled structures.
Bayesian-model-based methods are quite general, and thus, they are apt for a variety of material-discovery problems, such as identifying the low-energy region in a potential energy surface \cite{Toyoura2016}. 
For the interface structure, some studies \cite{Kiyohara2016,Kikuchi2017} proposed to apply Bayesian optimization to the GB-structure search, and its efficiency was confirmed, for example, by using the fcc-Cu $\Sigma5$ [001](210) CSL GB.
However, their search method is a standard Bayesian optimization method, i.e., same as the method in the case of bulk structures.
To our knowledge, a search methodology specifically for GBs has not been introduced so far.

As a general problem setting in the GB-structure search, we consider the exploration of a variety of rotation angles for a fixed rotation axis.
Suppose that we have $T$ different angles to search, and candidate structures are created by RBTs for each of them.
A naive approach to this problem is to apply some search method, such as Bayesian optimization \cite{Kiyohara2016,Kikuchi2017}, $T$ times separately. 
%
However, this approach is not efficient because it ignores the following two important characteristics of the GB structure:
\begin{enumerate}
 \item Energy-surface similarity: 
       The energy surfaces at different angles are often quite similar.
       This similarity is explained by the \emph{structural unit model} \cite{Sutton1983a, Sutton1983b, Sutton1983c, Sutton1990}, which has been widely accepted to describe GB structures in many materials.
       This model suggests that different GBs can contain common structural units, and that they share similar local atomic environments.
       Although structurally similar GBs can produce similar energy surfaces, the naive search does not utilize this similarity and restarts the structure search from scratch for each angle.
 \item Cost imbalance: 
       %
       %
       GB supercells usually have various sizes because of the 
variations in
the $\Sigma$ value, which is the inverse of the density of lattice sites. 
       This means that the computational cost for large $\Sigma$ GBs dramatically increases because the number of atoms in a supercell increases.              
       Thus, the structure search for large $\Sigma$ GBs is significantly more time-consuming than that for small $\Sigma$ GBs.
       For example, the computational time scale is $O(M) \sim O(M^{3})$ for $M$ number of atoms in the supercells, depending on the computational scheme.
\end{enumerate}
\figurename~\ref{fig:example-energy-surface} shows an example of this situation. 
The figure contains (a) an illustration of RBT and an atom removal from the boundary, (b) calculated stable GB energies, and (c) energy surfaces created by two-dimensional RBTs for the rotation angles $141^\circ$ (top), $134^\circ$ (bottom left) and $145^\circ$ (bottom right).
The entire landscape of the surfaces in \figurename~\ref{fig:example-energy-surface} (c) are similar, while their computational costs are significantly
different since the biggest supercell ($\Sigma89$) contains almost $10$ times larger number of atoms than the smallest supercell ($\Sigma9$).

\begin{figure}
 \begin{center}
  \includegraphics[clip,width=\textwidth]{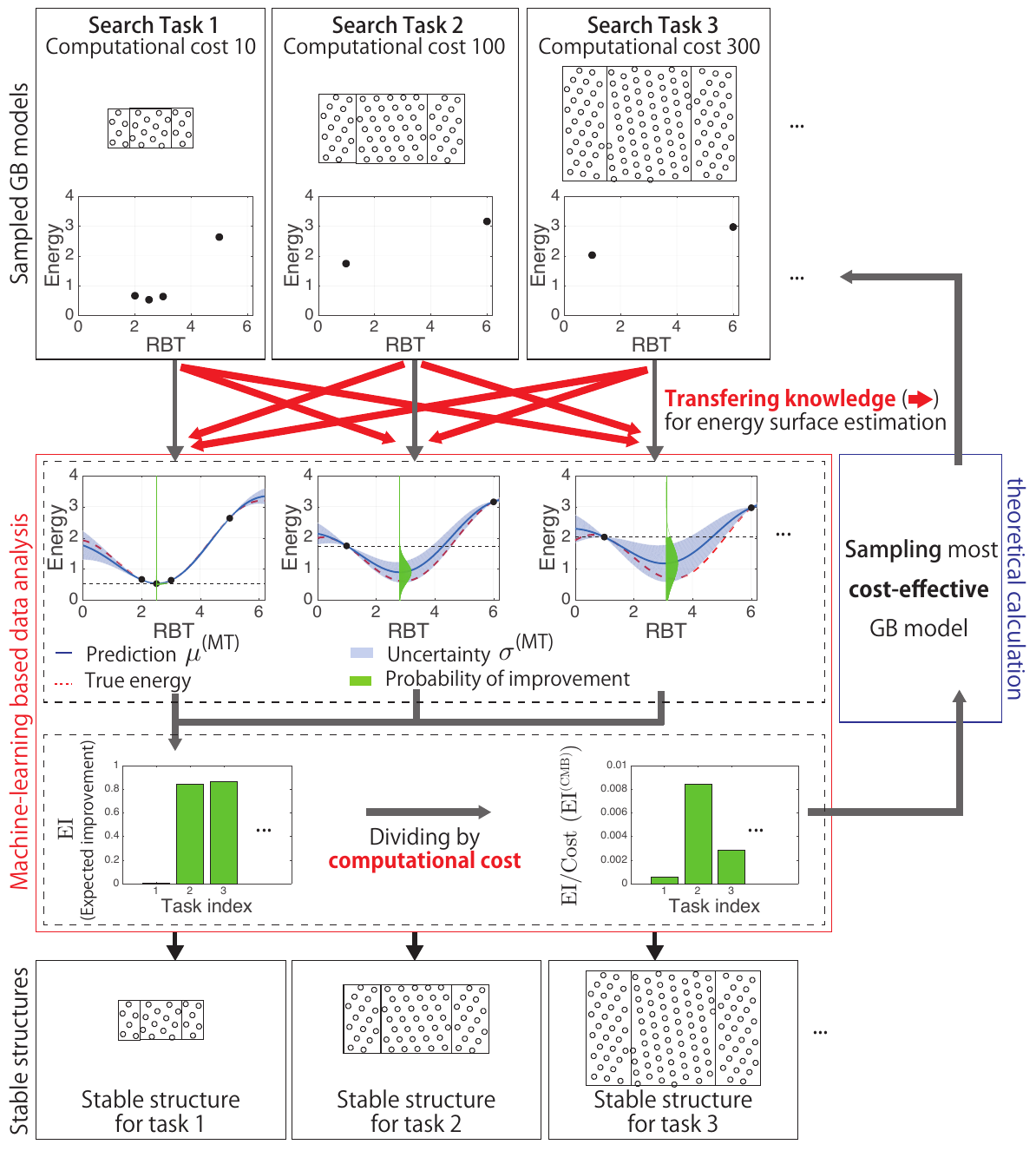}
  \caption{
  Schematic illustration of our proposed method.
  Our method transfers knowledge of observed GB models in different tasks (illustrated as red arrows).
  Machine learning constructs a probabilistic approximation of the energy surfaces based on information shared across the tasks, which results in better approximation accuracy compared to that realized by solving all tasks separately.
  %
  We also consider the cost discrepancy for given tasks by evaluating cost-effectiveness of each candidate GB model, which accelerates the search by reducing the number of model samplings for the high cost energy surfaces.
  }
  \label{fig:shematic-illust}
 \end{center} 
\end{figure}

In this paper, we propose a machine-learning-based stable structure search method that is particularly efficient for the GB-structure search.
Our proposed method, called cost-sensitive multi-task Bayesian optimization (CMB), takes the above two characteristics of GB structures into account.
For energy-surface similarity, we introduce a machine-learning concept called \emph{transfer learning} \cite{Pan2010}.
The basic idea of transfer learning is to transfer knowledge among different (but related) tasks to improve the efficiency of machine-learning methods.
In this study, a GB-structure search for a fixed angle is considered to be a ``task''.
When a set of tasks are similar to each other, information accumulated for one specific task can be useful for other tasks.
In our structure-search problem, a sampled GB model for an angle provides information for other angles because of the energy-surface similarity.
%
%
%
For the cost imbalance issue, we introduce a \emph{cost-sensitive} search.
Our method incorporates cost information into the sampling decision, which means that we evaluate each candidate based on both the possibility of an energy improvement and the cost of sampling.
By combining the cost-sensitive search with transfer learning, CMB accumulates information by sampling low cost surfaces in the initial stage of the search, and can identify the stable structures in high cost surfaces with a small number of sampling steps by using the transferred surface information.
\figurename~\ref{fig:shematic-illust} shows a schematic illustration of the entire procedure of CMB, which indicates that knowledge transfer, particularly from the low cost surfaces to the high cost surfaces, is beneficial for the structure search.
As a case study, we evaluate the cost-effectiveness of our method based on fcc-Al [110] tilt GBs: 
our proposed method determines
stable structures with $5$ mJ/m$^2$ average accuracy with only 
about $0.2$ \% of the computational cost of the exhaustive search.

\clearpage

\section*{Methods}
\label{sec:method}

\subsection*{Problem Setting}
\label{subsec:prob-setting}

GB energy is defined against the total energy of the bulk crystal as 
\begin{align}
 E_{\rm GB} = \frac{E_{\rm GB}^{\rm tot} - E_{\rm bulk}}{ 2 S },
\end{align}
where $E_{\rm GB}^{\rm tot}$ is the total energy of the GB supercell, $E_{\rm bulk}$ is the bulk energy with the same number of atoms as the GB supercell, and $S$ is the cross-section area of the GB model in the supercell.
In the denominator, the cross-section area $S$ is multiplied by $2$ since the supercell contains two GB planes as shown in \figurename~\ref{fig:example-energy-surface} (a) which is an example of a $\Sigma$9 GB model.
%
Note that the GB energy for each GB model is calculated through \emph{atomic relaxation}.

Suppose that we have $t = 1, \ldots, T$ different rotation angles $\theta_t$, for each of which we have $N_t$ candidate-GB models created by \emph{rigid body translations} (RBTs) with or without atom removal.
%
\figurename~\ref{fig:example-energy-surface} (a) also illustrates RBTs by which $N_t$ GB models are created.
%
The total number of the GB models is denoted as $N = \sum_{t = 1}^T N_t$.
We would like to search the stable GB structures with respect to \emph{all} of the given rotation angles.
A set of GB energies for all $N$ GB models is represented as a vector $\bE = (E_{\rm GB}^{(1)}, \ldots, E_{\rm GB}^{(N)})^\top$, where $E_{\rm GB}^{(i)}$ is the GB energy of the $i$-th GB model.

A stable structure search for some fixed angles can be mathematically formulated as a problem to find low energy structures with a smaller number of ``model sampling steps'' from candidates.
The number of candidate structures is often too large to exhaustively compute their energies, and we usually do not know the exact energy surface as a function in the search space.
This problem setting is thus called the \emph{black-box optimization problem} in the literature.
We call a stable structure search for each angle a ``task''.

Let $\tau_i \in \{ 1, \ldots, T \}$ be the task index that the $i$-th GB model is included, and $C_t$ be the cost to compute the GB energy in the $t$-th task.
%
We assume that the cost can be estimated based on the number of atoms $M$ in the supercell. 
For example, \emph{embedded atom method} (EAM) \cite{Mishin1999} with the cutoff radius needs $O(M)$ computations. 
Then, we can set $C_{t}$ as $M$.
Instead of counting the number of model samplings, we are interested in the sum of the cost $C_t$ of
the search process, for a
practical evaluation of the search efficiency.
Assuming that a set $\cS \subseteq \{ 1, \ldots, N \}$ is an index set of sampled GB models, the total cost of sampling is written as
\begin{align}
 C = \sum_{i \in \cS} C_{\tau_i}. 
 \label{eq:total-cost}
\end{align}

\subsection*{Knowledge-Transfer based Cost-Effective Search for GB Structures}

Our method is based on \emph{Bayesian optimization} which is a machine-learning-based method for solving general black-box optimization.
The basic idea is to estimate a stable structure iteratively, based on a probabilistic model that is statistically constructed by using already sampled structures.
\emph{Gaussian process regression} (GP) \cite{Rasmussen2005} is a probabilistic model usually employed in Bayesian optimization.
GP represents \emph{uncertainty} of unobserved energies by using a Gaussian random variable.
Let $\bx_i \in \RR^p$ be a $p$ dimensional descriptor vector for the $i$-th GB model, and $\bE_\cS$ be a energy vector for a set of sampled GB models.
The prediction of the $i$-th GB model is given by
\begin{align}
 f_i \mid \bE_{\cS} \sim \cN(\mu(\bx_i),\sigma(\bx_i)), 
 \label{eq:pred-dist-GP}
\end{align}
where $f_i \mid \bE_{\cS}$ is a random variable $f_i$ after observing $\bE_\cS$, and $\cN(\mu(\bx_i),\sigma(\bx_i))$ is a Gaussian distribution having $\mu(\bx_i)$ and $\sigma(\bx_i)$ as the mean and the standard deviation, respectively.
Bayesian optimization iteratively predicts the stable structure based on $\mu(\bx_i)$ and $\sigma(\bx_i)$.
See Supplementary Information 1 for details regarding the Bayesian optimization.

Although energy surfaces for different angles are often quite similar, simple Bayesian optimization cannot utilize such similarity.
In machine learning, it has been known that, for solving a set of similar tasks, transferring knowledge across the tasks can be effective.
This idea is called \emph{transfer learning} \cite{Pan2010}.
In particular, we introduce a concept called \emph{multi-task learning}, in which knowledge is transferred among multiple tasks, to accelerate convergence of multiple structure-search tasks of GB.

\begin{figure}
 \begin{center}
  \includegraphics[clip,width=\textwidth]{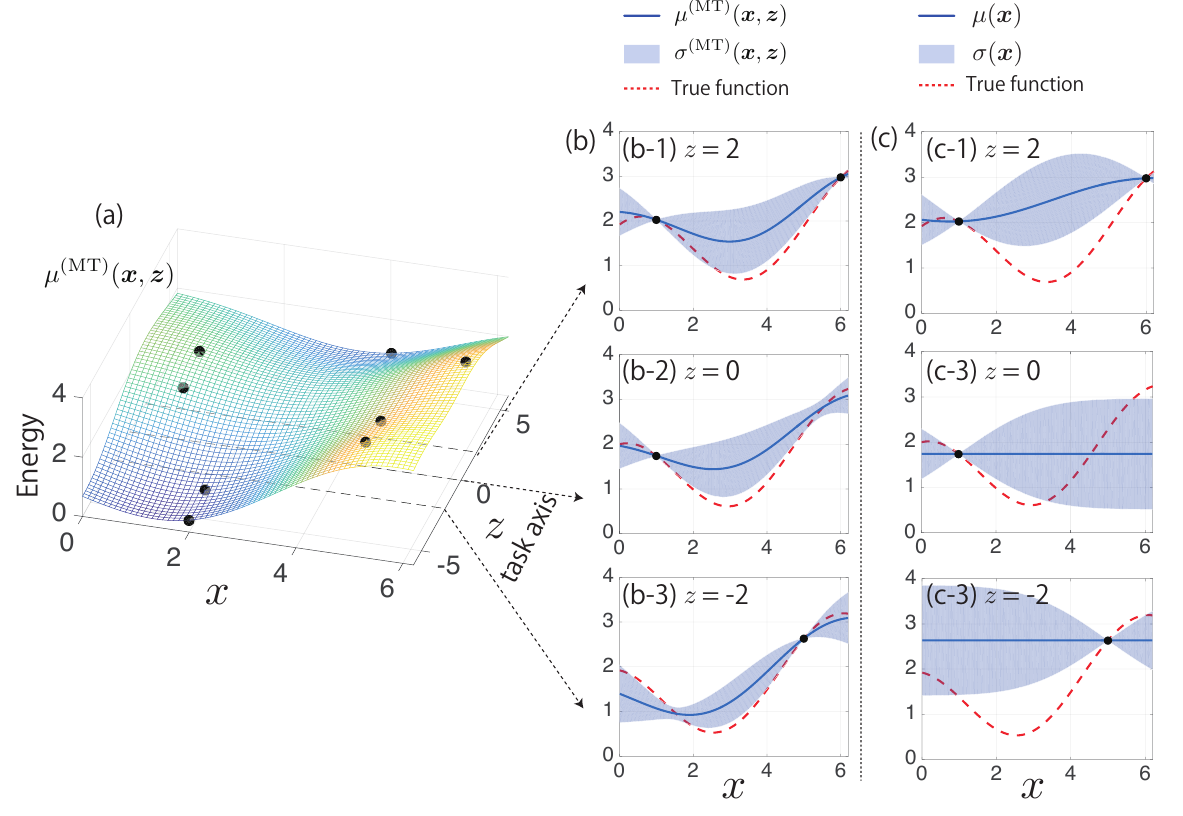}
  \caption{
  A schematic illustration of MGP. 
  The left three-dimensional plot (a) shows the MGP mean function surface $\mu^{\rm (MT)}$ in the joint space of the GB model descriptor $\bx$ and the task-specific descriptor $\bz$ (the black points are the sampled GB models).
  The center plots (b) show the surfaces for the three different tasks neighboring each other in the task axis.
  The blue lines are mean functions with the blue shaded standard deviations, and dashed red lines are underlying true functions.
  Since information on the sampled GB models is shared, all the mean functions of MGB provide better approximations for the true functions compared with the separated estimation of each task illustrated in (c).
  %
  %
  %
  %
  %
  }
  \label{fig:mtgp}
 \end{center}
\end{figure}

In addition to the structure descriptor $\bx$, we introduce a descriptor which represents a task.
Let $\bz_t \in \RR^q$ be a descriptor of the $t$-th task, called a task-specific descriptor, through which the similarity among tasks is measured. 
For example, a rotation angle can be a task-specific descriptor because surfaces for similar angles are often similar.
Hereafter, we refer to a descriptor $\bx$ as a structure-specific descriptor.
Given these two types of descriptors, we estimate the energy surface in the joint space of $\bx$ and $\bz$:
\begin{align}
 f_i \mid \bE_{\cS} \sim \cN(\mu^{\rm (MT)}(\bx_i, \bz_{\tau_i}), \sigma^{\rm (MT)}(\bx_i, \bz_{\tau_i})).
\end{align}
Here, the mean $\mu^{\rm (MT)}$ and standard deviation $\sigma^{\rm (MT)}$ are functions of both of the structure-specific descriptor $\bx$ and the task-specific descriptor $\bz$.
This model is called \emph{multi-task Gaussian process regression} (MGP)\cite{Bonilla2008}, and \figurename~\ref{fig:mtgp} shows a schematic illustration.
In the figure, 
information regarding the GP model is
transferred among tasks through ``task axis'', and it improves the accuracy of the surface approximation.
For a task-specific descriptor, we employed the rotation angle and radial distribution function in the later case study (See section ``GB Model and Descriptor'' for details).
Supplementary Information 2 provides for further mathematical details on MGP.

We propose combining MGP with Bayesian optimization, meaning that we determine the next structure to be sampled based on the probabilistic estimation of MGP.
%
Since knowledge transfer improves accuracy of GP (particularly for tasks in which there exists only a small number of sampled GB models), the efficiency of the search is also improved as illustrated in \figurename~\ref{fig:mtgp}.
After estimating the energy surface, Bayesian optimization calculates the \emph{acquisition function} using which we determine the structure to be sampled next.
A standard formulation of acquisition function is \emph{expected improvement} (EI) defined as the expectation of the energy decrease estimated by GP, which is also applicable in our multi-task GP case.
However, 
EI does not consider the cost discrepancy for the surfaces, which may necessitate a large number of sampling steps for high cost surfaces.
In other words, the total cost Eq.~(\ref{eq:total-cost}) is not taken into account by usual Bayesian optimization.

We further introduce a cost-sensitive acquisition function to solve this issue, and then the method is called \emph{cost-sensitive multi-task Bayesian optimization} (CMB).
%
%
To select the next candidate, each GB model is evaluated based not only on the possible decrease of the energy, but also on the computational cost of that GB model.
Our cost-sensitive acquisition function for the $i$-th GB model is defined by
\begin{align}
 {\rm EI}^{\rm (CMB)}_i = 
  \frac{
 {\rm EI}_i
  }{C_{\tau_i}},
 \label{eq:CSAF}
\end{align}
where ${\rm EI}_i$ is the usual expected improvement for the $i$-th GB model which purely evaluates the possible improvement.
%
This cost-sensitive acquisition function selects the best GB model to be sampled 
by considering
EI \emph{per} computational cost, while usual EI selects a structure 
by considering
the improvement in the energy decrease per sampling iteration.
%
%
%

\begin{figure}
 \begin{center}
  \includegraphics[clip,width=\textwidth]{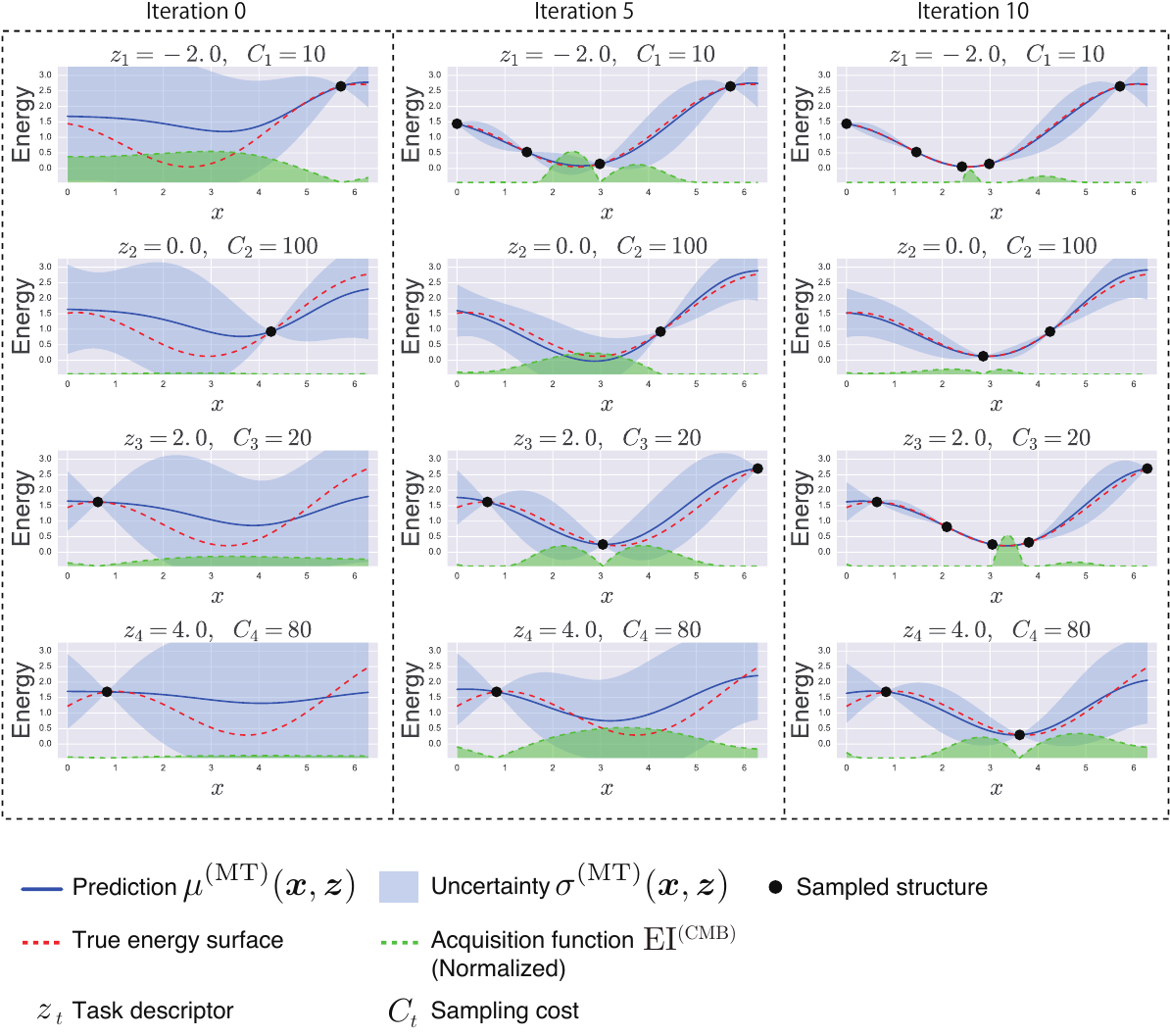}
  \caption{
  An illustrative example of the proposed method for the synthetic four tasks $t = 1, \ldots, 4$ with cost $C_1 = 10$, $C_2 = 100$, $C_3 = 20$, and $C_4 = 80$.
  In the iteration $0$, the initial points are randomly set.
  Our method first investigates the low cost surfaces $t = 1$ and $3$ as indicated by the acquisition function (green).
  In the iteration $5$, with the increase in the low cost surface points, uncertainty of the Gaussian process model is reduced even for the high cost surfaces $t = 2$ and $4$ in which no additional points are sampled yet. 
  Then, the acquisition function values for the high cost surfaces become relatively large because the possible energy improvement in the low cost surfaces is not significant compared to that in the Iteration 0. 
  %
  In the iteration $10$, the small energy points in the high cost surfaces are identified with a small number of model samplings.
  }
  \label{fig:CMTB-demo}
 \end{center}
\end{figure}

\figurename~\ref{fig:CMTB-demo} shows an illustrative demonstration of CMB.
In the figure, the two surfaces need low sampling costs and the other two surfaces need high sampling costs.
CMB first selects the low cost surfaces and accumulates surface information, using which the minimum energies for the high cost surfaces can be efficiently identified. 
This illustrates that CMB is effective for minimizing the GB energy with a small amount of the total cost Eq.~(\ref{eq:total-cost}).

\clearpage
\section*{Results (Case Study on fcc-Al)}
\label{sec:results}


\subsection*{GB Model and Descriptor}
\label{sec:GB-model-setting}

\begin{figure}
 \begin{center}
  \includegraphics[clip,width=0.8\textwidth]{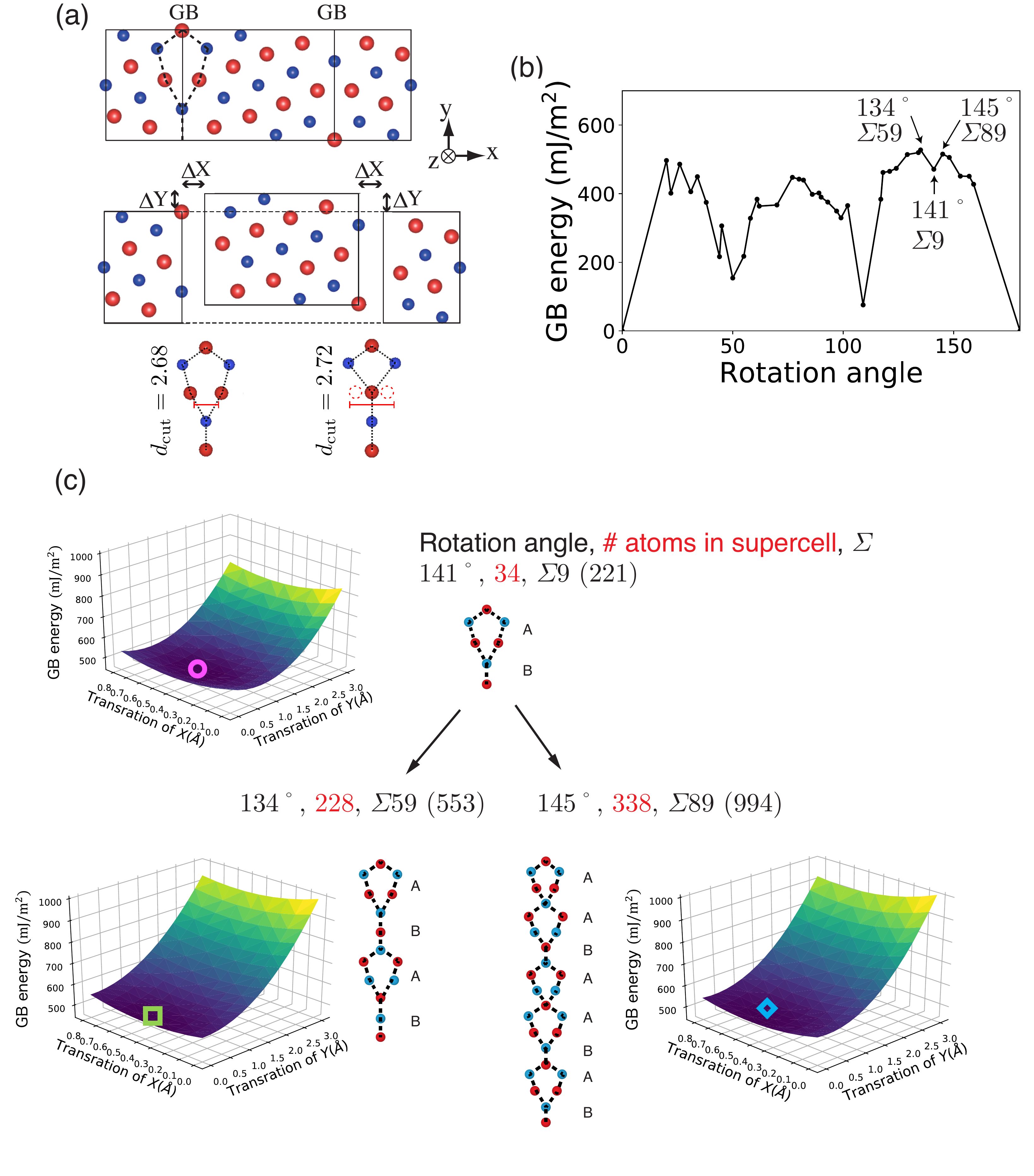} 
  \caption{
  (a)
  Atomic configuration of the GB supercell of the fcc-Al [110] $\Sigma9$ symmetric tilt GB. 
  The red and blue balls denote Al atoms in the (110) and (220) atomic layers.
  Based on this GB model, four microscopic parameters are optimized.
  One of two grains is rigidly shifted with $\Delta X$, $\Delta Y$ and $\Delta Z$, and an atomic pair within the cutoff distance $d_{\rm cut}$ is replaced with a single atom located at the center of the original pair.
  (b) Calculated stable GB energies as a function of the rotation angle for fcc-Al [110] tilt GB. 
  (c) Energy surfaces created by 
  RBTs for the angles $141^\circ$ (top), $134^\circ$ (bottom left), and $145^\circ$ (bottom right).
  For illustrative purpose, we here only show two-dimensional RBTs on $X$ and $Y$.
  The structural units are also shown along with the surfaces in which the units are denoted as \texttt{A} and \texttt{B}.
  The red and blue balls denote Al atoms in (110) and (220) atomic layers.
  The markers on the surfaces indicate their minimums.
  The numbers of atoms in the supercells, which determine computational cost, are written in red.
  } \label{fig:example-energy-surface}
 \end{center}
\end{figure}

We first constructed fcc-Al [110] symmetric tilt (ST) GBs using the coincidence site lattice (CSL) model. 
The CSL is usually characterized by the $\Sigma$ value, which is defined as the reciprocal of the density of the coincident sites. 
%
\figurename~\ref{fig:example-energy-surface} (a) shows an example of a supercell of a $\Sigma9$ STGB model.
Two symmetric GBs are introduced to satisfy three-dimensional periodicity. 
To avoid artificial interactions between GBs, we set the distances between GBs to more than 10 \AA.
%
For the energy calculations and atomic relaxations, we used the EAM potential for Al in Ref. \cite{Mishin1999}, and the computational time scales as $O(M)$ for the number of atoms $M$ with the linked-list cell algorithm.
%
\figurename~\ref{fig:example-energy-surface} (a) also shows the construction of a supercell by RBT from the STGB model.
%
%
{
GB models contain largely different numbers of atoms in the supercells from 
$36$
to 
$388$
which results in a strong cost imbalance in the search space.  
The number of atoms $M$ for all 
$38$
angles are shown in Supplementary Information 3.
For each angle, the {three-dimensional RBTs, denoted as $\Delta X$, $\Delta Y$ and $\Delta Z$} which are illustrated in 
\figurename~\ref{fig:example-energy-surface} (a), were generated.
%
The grid space is 
$0.1 {\rm \AA}$ 
for the direction $\Delta X$, 
$0.2 {\rm \AA}$ 
for the direction $\Delta Y$,
and
$0.1 {\rm \AA}$ 
for the direction $\Delta Z$.
%
In atomic columns, if the two atoms in an atomic pair are closer to each other than the cut-off distance, one atom from the pair is removed
More precisely, an atomic pair within the cutoff distance is replaced with a single atom located at the center of the original pair.
In this study, the cut-off distance 
is varied between 1.43 and 2.72 \AA, i.e. 0.5 and 0.95 times the equilibrium atomic distance, respectively. 
For example, two models for $\Sigma 9 $, where an atomic pair is replaced or not replaced, can be considered as illustrated in \figurename~\ref{fig:example-energy-surface} (a).
In total, we created 
$157680$
candidate GB models for which the exhaustive search is computationally quite expensive.
}
%


As the structure-specific descriptor for each GB model $\bx$, we employed the {three-dimensional axes of RBTs: $\Delta X$, $\Delta Y$, and $\Delta Z$}.
%
%
For the task-specific descriptor $\bz$, we used the rotation angle $\theta$ and radial distribution function (RDF) of the 
{$(\Delta X,\Delta Y,\Delta Z) = (0,0,0)$}
GB model.
As an angle descriptor, we applied the following transformation to the rotation angles: $\tilde{\theta}_t = \theta_t$ if $\theta_t \leq 90$, otherwise $\tilde{\theta}_t = 180 - \theta_t$. 
In the case of the fcc-Al [100] GBs, $\theta_t$ and $\tilde{\theta}_t$ are equivalent.
Although the complete equivalence does not hold for fcc-Al [110] GBs, we used this transformed angle as an approximated similarity measure.
For the RDF descriptor, we created a $100$-dimensional vector $\brho \in \RR^{100}$ by taking $100$ equally spaced grids from $0$ to $6$ \AA.
The task-specific descriptor is thus written as $\bz_t = (\tilde{\theta}_t, \brho_t^\top)^\top$.
%
In other words, two tasks which have similar angles and RDFs simultaneously are regarded as similar in MGP.
%
%
The cost parameter $C_t$ was set by the number of atoms in each supercell.
Detail of the parameter setting of Bayesian optimization is shown in Supplementary Information 4.

\subsection*{Performance Evaluation}
\label{subsec:performance}

To validate the effectiveness of our proposed method, we compared the following four methods (methods 3 and 4 are newly proposed in this paper.):
\begin{enumerate}
 \item random sampling (Random): At each iteration, the next candidate was randomly selected with uniform sampling.
 \item single task Bayesian optimization (SB): SB is the usual Bayesian optimization for a single task.
       At each iteration, a GB model which had the maximum EI was selected across all the angles.
       %
       %
 \item multi-task Bayesian optimization (MB): MB is Bayesian optimization with multi-task GP in which knowledge of the energy surfaces is transferred to different angles each other.        
       The acquisition function is the usual EI.
 \item cost-sensitive multi-task Bayesian optimization (CMB): CMB is MB with the cost-sensitive acquisition function defined by Eq.~(\ref{eq:CSAF}).
\end{enumerate}
All methods start with one randomly selected structure for each angle.

\begin{figure}
 \begin{center}
  \includegraphics[clip,width=0.8\textwidth]{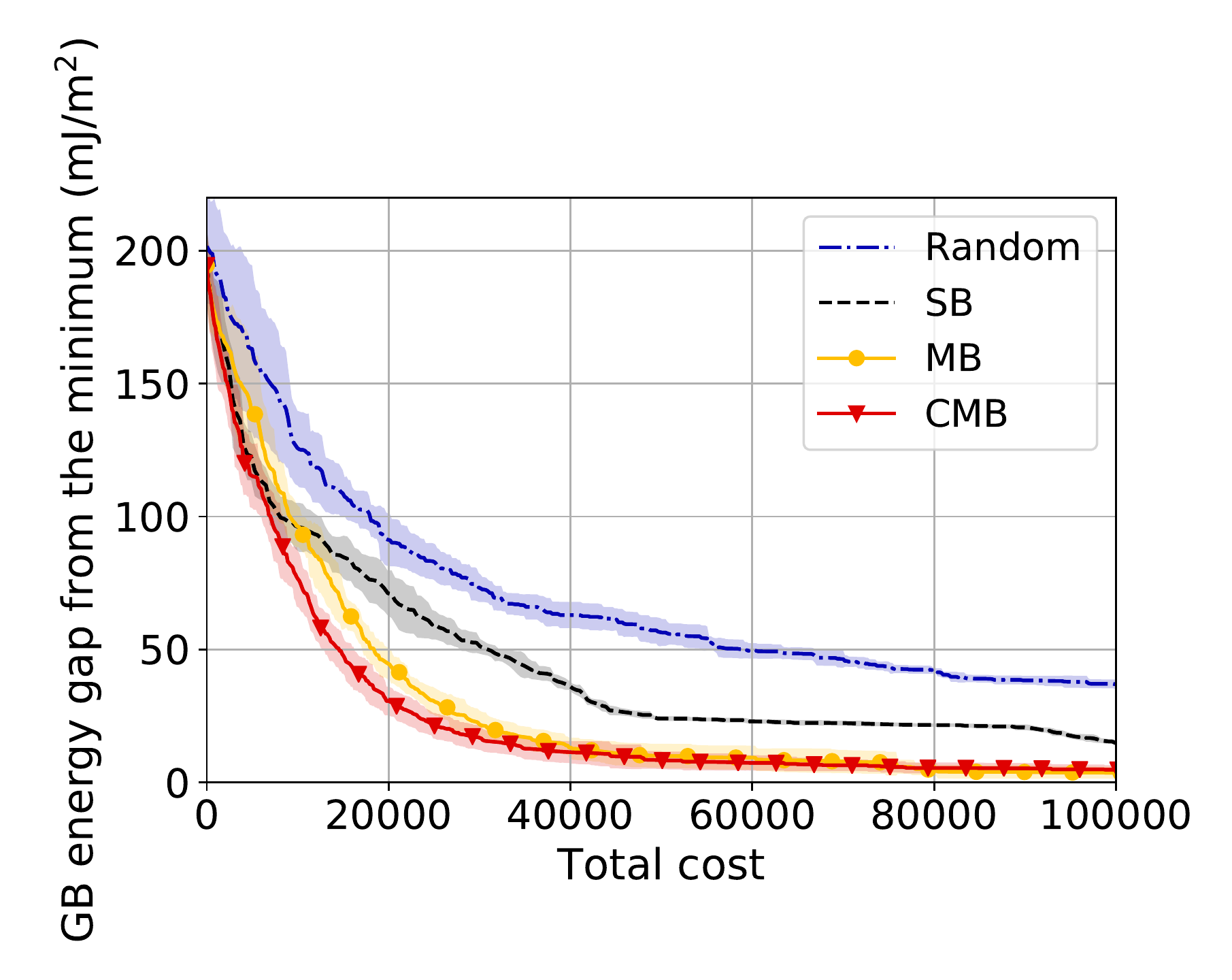}
  \caption{
  The GB energy gaps from the minimums to the identified structure by each method.
  The vertical axis is the mean for all angles.
  The shaded region represents the standard deviation for five runs.
  }
  \label{fig:error-transition}
 \end{center}
\end{figure}

\figurename~\ref{fig:error-transition} shows the results.
We refer to the difference between the lowest energy identified by each search and the true minimum as an energy gap.
The vertical axis of the figure is the average of the gaps for the 
{$38$}
different angles, and the horizontal axis is the total cost (\ref{eq:total-cost}).
All values are averages of {$5$} trials with different initial structures.
%
%

{We first see that CMB has smaller energy than the other three methods.}
%
Focusing on the difference between the single-task method and the multi-task-based methods, we see that the convergence of SB is much slower than that of multi-task based methods (MB and CMB).
%
%
%
We also see that the cost-sensitive search improved the convergence
(Note that although the cost-sensitive search is applicable to SB, it is not essentially beneficial because SB does not transfer information accumulated for low cost surfaces to high cost surfaces.).

\begin{figure}
 \begin{center}  
  \includegraphics[clip,width=0.8\textwidth]{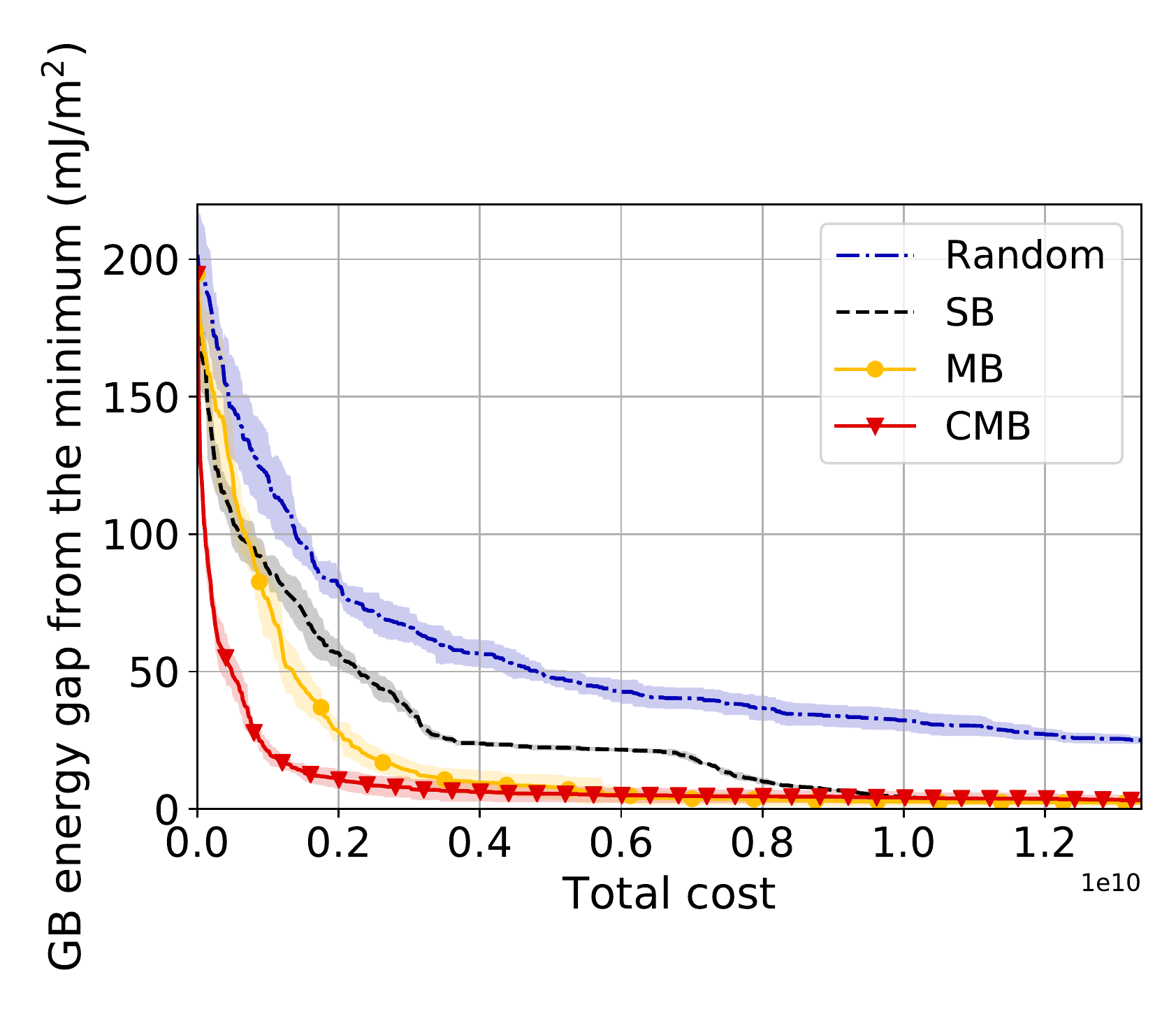}
  \caption{
  The GB energy gaps for the $O(M^3)$ cost setting.
  }
  \label{fig:error-transition-c3}
 \end{center}
\end{figure}

{
To validate the effectiveness of our approach in a more computationally expensive setting, we consider the case that $O(M^3)$ computations are necessary for the atomic relaxation.
By setting the cost parameter $C_t$ as the cube of the number of atoms (i.e., $M^3$), we virtually emulated this situation with the same dataset.
\figurename~\ref{fig:error-transition-c3} shows the energy gap.
Here, the horizontal axis is the sum of the cube of the number of atoms $M^3$ for the calculated GB models.
Same as \figurename~\ref{fig:error-transition}, MB and CMB show better performance than the naive SB.
In particular, CMB rapidly decreased the energy gap than the other methods.
Because of the larger sampling cost, the cost-sensitive strategy showed a greater effect on the search efficiency.
}

\section*{Discussion}

\begin{figure}
 \begin{center}
  \includegraphics[clip,width=0.8\textwidth]{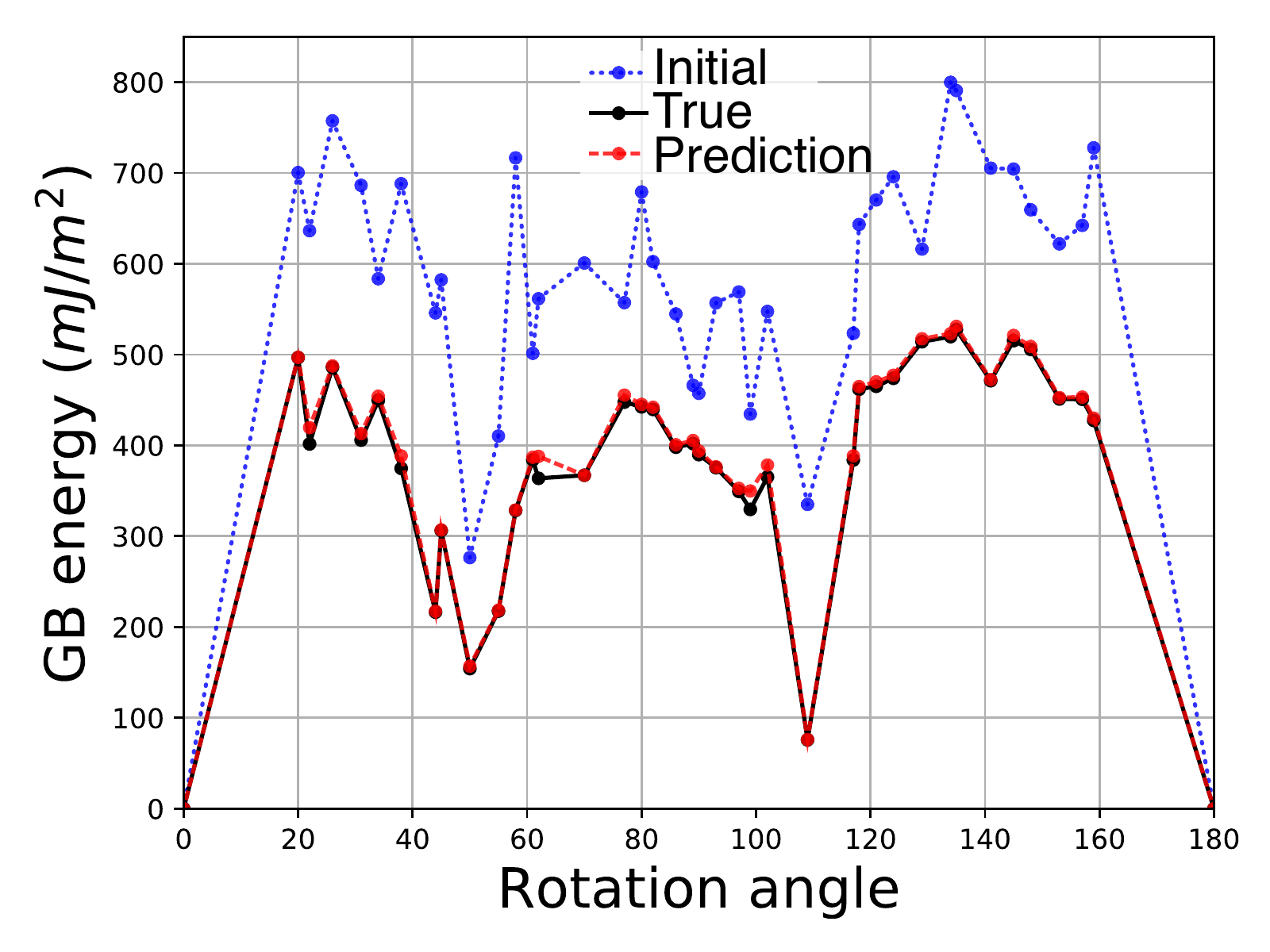}
  \caption{
  Stable GB energy as a function of the rotation angle.
  The solid line represents the true stable energy obtained by computing all GB models exhaustively.
  %
  %
  The dotted line corresponds to the average energy at the initial step of Bayesian optimization.
  The dashed line represents the average energy obtain by CMB with the cost value of 100000, which is about 0.3\% ($\approx 100000/33458160$) of the cost of the exhaustive search.
  }
  \label{fig:cusp}
 \end{center}
\end{figure}

The acceleration of the structure search is essential for material discovery in which a huge number of candidate structures are needed to be investigated.
In our case study using the fcc-Al [110] tilt GBs, the sum of the computational cost $C_t$ for all candidate structures is $\sum_{i=1}^N C_{\tau_i} = 33458160$ when $C_t$ is set as per the $M$ (i.e., $O(M)$ setting). 
%
The total computational cost that CMB needed to reach 
the average energy gaps
$10$ mJ/m$^2$ and
$5$ mJ/m$^2$ 
were
$0.001 \approx 43891.8 / 33458160.0$ 
and 
$0.002 \approx 76937.6 / 33458160.0$, 
respectively.
%
In other words, with only about 0.2 \% of the computation steps of the exhaustive search, CMB achieved 5 mJ/m$^2$ accuracy.
Figure~\ref{fig:cusp} compares the energy between the true stable structure and the structure identified by CMB, which shows that our method accurately identified the dependency of energy on the angle, with a low computational cost.

To summarize, we have developed a cost-effective simultaneous search method for GB structures based on two machine-learning concepts: transfer learning and cost-sensitive search.
Since amount of data is a key factor for data-driven search algorithms, knowledge transfer, by which data is shared across different tasks, is an important technique to accelerate the structure search.
Although the concept of multi-task learning is widely accepted in the machine-learning community, our method is the first study which utilizes it for fast exploration of stable structures.
%
Our other contribution is to introduce the concept of the cost-sensitive evaluation into the structure search.
For efficient exploration, the diversity of computational cost should be considered, though this issue has not been addressed in the context of the structure search.
Although we used the EAM potential as an example, the cost-imbalance issue would be more severe for computationally more expensive calculations such as density functional theory (DFT) calculations.

\section*{Data availability}

The gain-boundary structure data and our Bayesian optimization code are available on request.

\section*{Acknowledgement}

We would like to thank R. Arakawa for helpful discussions on GB models.
This work was financially supported by grants from the Japanese Ministry of Education, Culture, Sports, Science and Technology awarded to I.T. (16H06538, 17H00758) and M.K. (16H06538, 17H04694); from Japan Science and Technology Agency (JST) PRESTO awarded to M.K. (Grant Number JPMJPR15N2); and from the ``Materials Research by Information Integration'' Initiative (MI$^2$I) project of the Support Program for Starting Up Innovation Hub from JST awarded to T.T., I.T., and M.K.

\section*{Author contributions}

T.Y. implemented all machine learning methods. T.T. constructed the grain-boundary database, and contributed to writing the manuscript. I.T. conceived the concept and contributed to writing the manuscript. M.K. conceived the concept, designed the research, and wrote the manuscript.

\bibliographystyle{apsrev}
\bibliography{ref}

\clearpage



\clearpage

\clearpage

\clearpage

\clearpage

\clearpage

\clearpage

\clearpage

\clearpage



\section*{Supplementary Information 1}

Here, we briefly review a basic concept and technical details of Bayesian optimization for some fixed $t$-th angle, which we call single-task Bayesian optimization (SB) in this study.

%
GP represents \emph{uncertainty} of unobserved energies by using a random variable vector with a multivariate Gaussian distribution:
\begin{align}
 \*f \sim \cN(\bu,\bK),
\end{align}
where $\*f = (f_1, \ldots, f_{N_t})^\top$ is a vector of random variables for approximating the energies $\bE$, and $\cN(\bu,\bK)$ is a Gaussian distribution having $\bu \in \RR^{N_t}$ as the mean vector and $\bK \in \RR^{N_t \times N_t}$ as the covariance matrix.
Note that since we only focus on the search for a fixed angle $\theta_t$, the indexes of GB models are $1, \ldots, N_t$.
Let $\bx_i \in \RR^p$ be a $p$ dimensional descriptor vector for the $i$-th GB model.
The $i,j$-th element of the covariance matrix is defined by a \emph{kernel function} $k(\bx_i,\bx_j)$ which gives the similarity between two arbitrary GB models $i$ and $j$.
As a kernel function $k: \RR^p \times \RR^p \rightarrow \RR$, the following Gaussian kernel is often employed:
\begin{align}
 k(\bx_i,\bx_j) = \exp \left( - \gamma \| \bx_i - \bx_j \|^2_2 \right),
\end{align}
where $\gamma > 0$ is a scaling parameter and $\| \cdot \|_2$ is the $L_2$ norm.

When we already have GB energies for a subset of GB models $\cS \subseteq \{ 1, \ldots, N_t \}$, GP updates its predictions for unknown GB energies using a conditional probability.
Let $\bv_\cS$ be a sub-vector of an arbitrary vector $\bv \in \RR^{N_t}$ with the elements corresponding to $\cS$, and $\bM_{\cS}$ be a sub-matrix of an arbitrary matrix $\bM \in \RR^{N_t \times N_t}$ with the rows and the columns corresponding to $\cS$.
The conditional probability, called \emph{predictive distribution}, of the $i$-th GB model given energies for $\cS$ is written as
\begin{align}
 f_i \mid \bE_{\cS} \sim \cN(\mu(\bx_i),\sigma(\bx_i)), 
 \label{eq:pred-dist-GP}
\end{align}
where $f_i \mid \bE_{\cS}$ is a random variable $f_i$ after observing $\bE_\cS$, and 
\begin{align}
  \mu(\bx_i) &= \bK_{i,\cS} \left( \bK_{\cS} + \epsilon \bI \right)^{-1} (\bE_\cS - \bu_\cS),\\
  \sigma(\bx_i) &= k(\bx_i,\bx_i) - \bK_{i,\cS} \left( \bK_{\cS} + \epsilon \bI \right)^{-1} \bK_{\cS,i}, 
\end{align}
in which $\bK_{i,\cS}$ is a row vector having the $i$-th row and the columns of $\cS$ in $\bK$ (and $\bK_{\cS,i}$ is its transpose), and $\epsilon \geq 0$ is a noise term.

In Bayesian optimization, a function called \emph{acquisition function} evaluates a possibility that each candidate GB model would be more stable than the sampled GB models in $\cS$.
\emph{Expected improvement} (EI) is one of most standard acquisition functions to select the next structure:
\begin{align}
 {\rm EI}_i = \EE_{f_i \mid \bE_{\cS}}
 \left(
 \max\{ 0, E^{\rm best} - f_i \}
 \right), \text{ for } i = 1, \ldots, N_t,
 \label{eq:EI}
\end{align}
where $\EE_{f_i \mid \bE_{\cS}}$ is an expectation with respect to $f_i \mid \bE_{\cS}$, and $E^{\rm best}$ is the minimum energy among already computed GB models $\cS$.
EI is the expected value (based on the predictive distribution of the current GP) of the energy decrease.
Bayesian optimization iteratively selects a next GB model by taking the maximum of EI, and the newly computed GB model is added to $\cS$.
Even if we have multiple energy surfaces, we can apply GP separately, and choose the maximum of EI among all surfaces as the next candidate.

\clearpage

\section*{Supplementary Information 2}

To transfer knowledge among different tasks, we employ \emph{multi-task Gaussian process regression} (MGP) \cite{Bonilla2008}.
In addition to the structure descriptor $\bx$, MGP introduces a descriptor which represents a task.
Let $\bz_t \in \RR^q$ be a descriptor of the $t$-th task, and $k^{\rm (task)}: \RR^q \times \RR^q \rightarrow \RR$ be a kernel function for a given pair of tasks.
%
%
The task kernel function $k^{\rm (task)}(\bz_t,\bz_{t'})$ provides the similarity of two given tasks $t$ and $t'$.
We employ the following form of the kernel function to define $k^{\rm (task)}$:
\begin{align}
 k^{\rm (task)}(\bz_t,\bz_{t'}) = 
 \alpha \exp \left( - \gamma^{\rm (task)} \| \bz_t - \bz_{t'} \|_2^2 \right) 
 + (1 - \alpha) \delta_{t,t'}
\end{align}
where $\gamma^{\rm (task)} > 0$ and $\alpha \in [0,1]$ are parameters, and $\delta_{t,t'}$ is defined as $1$ if $t = t'$; otherwise, it is $0$.
The additional parameter $\alpha$ is to control the independence of tasks.
When we set $\alpha = 1$, $k^{\rm (task)}(\bz_t,\bz_{t'})$ is $1$ only when $t = t'$, and is $0$ for all other cases.
This special case is reduced to apply $T$ separated Gaussian process models for all tasks independently.

MGP is defined by a kernel constructed as a product of two kernels on $\bx$ and $\bz$.
The $i,j$-element of the covariance matrix of MGP is defined as
\begin{align}
 \bK_{i,j}^{\rm (MT)} = k(\bx_i,\bx_j) k^{\rm (task)}(\bz_{\tau_i},\bz_{\tau_j}), \text{ for }
 i,j = 1, \ldots, N.
\end{align}
Note that we use the index $1, \ldots, N$ across all $T$ tasks, and the task in which the $i$-th GB model is contained is represented as $\tau_i \in \{ 1, \ldots, T \}$.
Then, we define random variables for unobserved energies as follows: 
\begin{align}
 \*f \sim \cN(\bu,\bK^{\rm (MT)}),
\end{align}
where $\*f = (f_1, \ldots, f_N)^\top$ is the random variable vector for GB energy and $\bu \in \RR^N$ is a mean vector.
Given a set of already sampled GB models $\cS \in \{ 1, \ldots, N \}$, the predictive distribution is derived by the same manner as in GP:
\begin{align}
 f_i \mid \bE_{\cS} \sim \cN(\mu^{\rm (MT)}(\bx_i, \bz_{\tau_i}), \sigma^{\rm (MT)}(\bx_i, \bz_{\tau_i})), 
 \label{eq:pred-dist-MTGP}
\end{align}
where 
\begin{align}
  \mu^{\rm (MT)}(\bx_i, \bz_{\tau_i}) &= \bK^{\rm (MT)}_{i,\cS} \left( \bK^{\rm (MT)}_{\cS} + \epsilon \bI \right)^{-1} (\bE_\cS - \bu_\cS),\\
  \sigma^{\rm (MT)}(\bx_i , \bz_{\tau_i}) &= \bK^{\rm (MT)}_{i,i} - \bK^{\rm (MT)}_{i,\cS} \left( \bK^{\rm (MT)}_{\cS} + \epsilon \bI \right)^{-1} \bK^{\rm (MT)}_{\cS,i}.
\end{align}
The only difference in GP and MGP is in their kernel matrices $\bK$ and $\bK^{\rm (MT)}$.
Unlike usual GP kernel $\bK$, the MGP kernel $\bK^{\rm (MT)}$ contains similarity between tasks and thus it can transfer information among different tasks.

\clearpage

\section*{Supplementary Information 3}

\begin{figure}[h]
 \begin{center}
  \includegraphics[clip,width=0.8\textwidth]{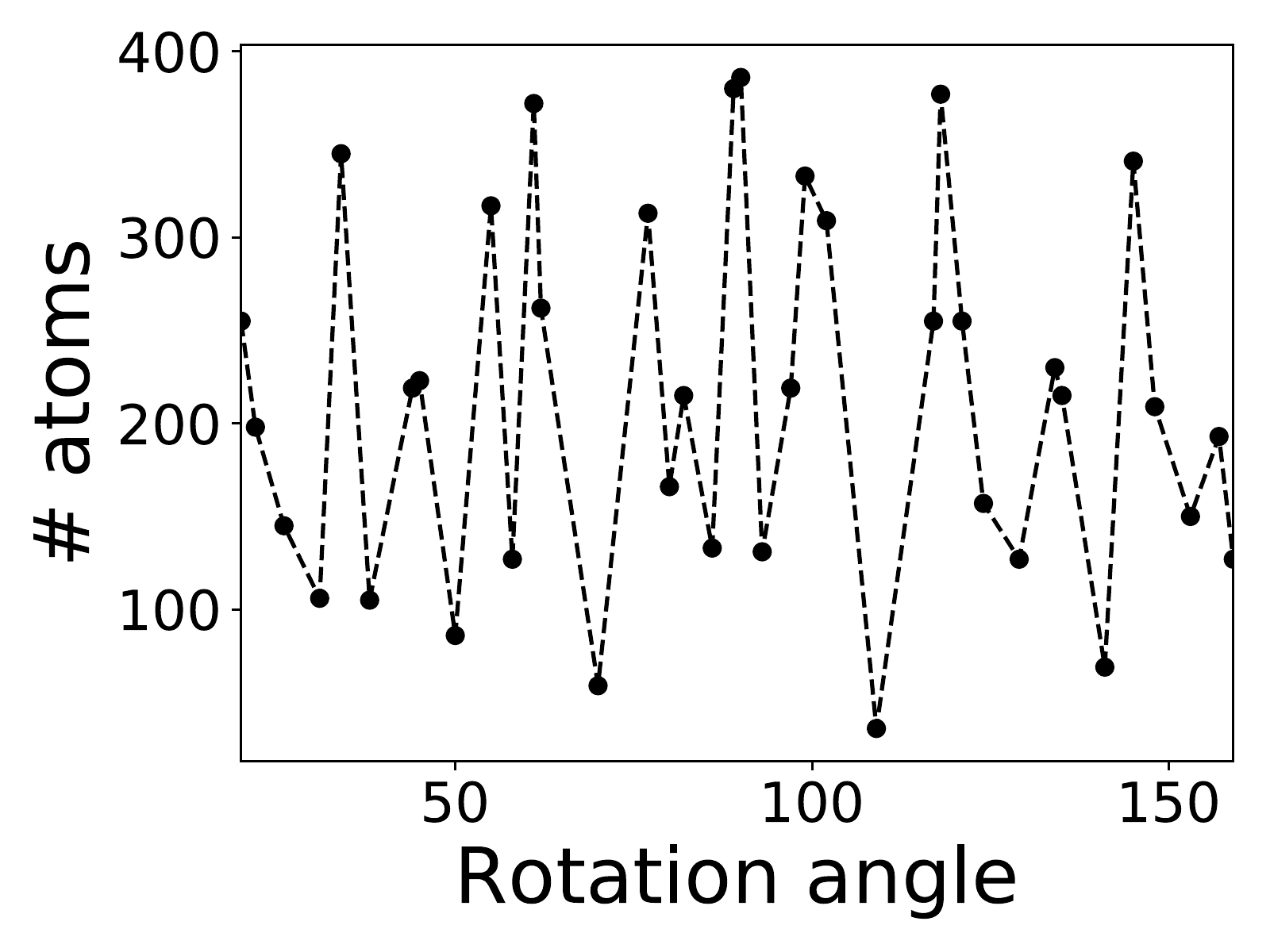}
  \caption{The number of atoms contained in the supercells of the different angles.}
  \label{fig:cost-plot}
 \end{center}
\end{figure}

\figurename~\ref{fig:cost-plot} shows the number of atoms in our GB dataset.

\clearpage

\section*{Supplementary Information 4}

The kernel function (See Supplementary Information 1 and 2 for definition of kernel) for $\bx$ was the Gaussian kernel with the parameter $\gamma$ set by median heuristics ($\gamma$ is set as the reciprocal of median of the squared distances).
The task kernel is defined as
\begin{align*}
 k^{\rm (task)}(\bz_t,\bz_{t'}) = 
 \alpha 
 \exp \left( - \gamma_\theta^{\rm (task)} \| \tilde{\theta}_t - \tilde{\theta}_{t'} \|_2^2 
 - \gamma_{\rm RDF}^{\rm (task)} \| \brho_t - \brho_{t'} \|_2^2 \right) 
 + (1 - \alpha) \delta_{t,t'} 
\end{align*}
where $\brho \in \RR^{100}$ is a vector created from RDF by taking $100$ equally spaced grids from $0$ to $6$ \AA.
In this case, the task-specific descriptor is written as $\bz_t = (\tilde{\theta}_t, \brho_t^\top)^\top$, and the kernel evaluates the task similarity based on both of the angle and RDF.
In other words, two tasks which have similar angles and RDFs simultaneously are regarded as similar in MGP.
The parameters $\gamma_\theta^{\rm (task)}$ and $\gamma^{\rm (task)}_{\rm RDF}$ were set by median heuristics again, and the independency parameter $\alpha$ is estimated by \emph{marginal likelihood maximization} \cite{Rasmussen2005}.
For each task, the values of the mean parameter $\bu$ was set separately as the average of sampled GB energies.
The noise term parameter $\epsilon$ 
was also selected by marginal likelihood maximization.
The parameter tuning for $\alpha$ and $\epsilon$ was performed every $10$ samplings.

\end{document}